\documentclass[12pt]{iopart}
\usepackage{iopams} 
\usepackage{hyperref}
\usepackage{siunitx}
\usepackage{graphicx}
\usepackage{multirow}

\usepackage{xcolor}
\newcommand{\sbd}[1]{\textcolor{red}{#1}}

\newcommand{\gtf}[1]{\textcolor{cyan}{#1}}

\renewcommand{\etal}{\textit{et al.}}

\begin{document}

\title[Dye Attenuation Without Dye]{Dye Attenuation Without Dye: Quantifying Concentration Fields with Short-wave Infrared Imaging}

\author{G. T. Fortune$^1$, M. A. Etzold$^{1,2}$, J. R. Landel$^{3,4}$ and Stuart B. Dalziel$^1$}

\address{$^1$ Department of Applied Mathematics and Theoretical Physics, Centre for Mathematical Sciences, University of Cambridge, Wilberforce Road, Cambridge CB3 0WA, United Kingdom}
\address{$^2$ The Defence Science and Technology Laboratory, Porton Down, B383B Room 1/6, Salisbury, Wiltshire, SP4 0JQ}
\address{$^3$ Department of Mathematics, Alan Turing Building, University of Manchester, Oxford Road, Manchester, M13 9PL, UK}
\address{$^4$ Universite Claude Bernard Lyon 1, Laboratoire de Mecanique des Fluides et d'Acoustique (LMFA), UMR5509, CNRS, Ecole Centrale de Lyon, INSA Lyon, 69622
Villeurbanne, France}

\ead{gtf22@cam.ac.uk}
\vspace{10pt}

\begin{abstract}

Dye attenuation, or photometric imaging, is an optical technique commonly used in fluid dynamics to measure tracer concentration fields and fluid thicknesses under the assumption that the motion of the dye is representative of the fluid motion and that its presence does not affect the behaviour of the system. However, in some systems, particularly living biological systems or those with strong chemical interactions and reactions, the addition of dye may non-trivially influence the system and may not follow the fluid containing it.

To overcome this, we demonstrate how short-wave infrared imaging can be used to measure concentration and height profiles of water and other liquids without the introduction of dye for heights down to \SI{0.2}{\milli\metre} with spatial and temporal resolutions of the order of \SI{50}{\micro\metre} per pixel and $120$ fps respectively. We showcase the utility of this technique by demonstrating its ability to accurately track the temporal evolution of the total water content of two model systems, namely a water drop spreading on a glass slide and spreading within a hydrogel sheet, validating both against an analytical mass balance. 
Finally, we discuss how the spectral resolution of the present setup could be increased to the point that concentrations within a multi-component system containing more than one type of liquid could be quantified.
\end{abstract}

\noindent\submitto{\MST}

\section{Introduction}
As light passes through a dyed fluid, its intensity is diminished. After calibration, the attenuation of light can then be related to the dye concentration integrated along the light path or the thickness of the fluid layer. This technique is known in fluid dynamics as \emph{dye attenuation} \cite{Cenedese98}. Such measurements have over the years been used extensively in a range of fluid dynamics contexts, including mixing in turbulent jets \cite{landel12}, mixing in stratified flows \cite{davies_wykes_dalziel_2014, Allgayer12}, rotating fluid systems \cite{Holford96} and gravity currents \cite{Hacker96}. The applicability of dye attenuation in these systems relies on the assumption that the motion of the dye is dependent only on the motion of the liquid, namely it is a passive tracer whose transport is dominated by the motion of the liquid medium.

Dye attenuation is also of interest to a wide range of fluid-dynamics experiments in which these assumptions do not hold. For example, Landel \etal{} \cite{Landel16} considered the removal of a polymer-thickened contaminant in a shear flow. Since transport within the droplet is diffusive, the dye transport is independent of the bulk fluid of the contaminant. Kim \etal{} \cite{Kim03} showed that adding vegetable dye to distilled water fundamentally altered the rheological properties of the resulting non-Newtonian dye-water solution for high enough dye concentrations (greater than $2\%$ by volume). Furthermore, dyes can often change both the density of the liquid and surface tension \cite{Zhang19} at fluid-fluid interfaces, affecting buoyancy driven flows and free-surface flows respectively. 

Absorbing polymer systems -- such as hydrogels -- have recently received considerable interest in the fluids community \cite{Bertrand16,Engelsberg:2013,Etzold21,Hennessy21,Butler22}. They exhibit swelling due to strong chemical interactions between the solvent (typically water) and the polymer. Theoretical poroelastic frameworks can be used to make significant analytical progress \cite{Fortune21,Etzold22}. However, these systems cannot simply be treated as two-phase porous media; rather, they must be considered as a non-ideal single-phase mixture where a significant contribution to its physico-chemical properties arises from inter-molecular forces \cite{Doi:2009}. The addition of a dye considerably increases the complexity of these systems through modification of these inter-molecular forces. The dye motion may be different to the water motion. Furthermore, in this three-component mixture, phenomena such as miscibility gaps may occur (i.e. combinations where all three components cannot coexist as a single-phase solution but must instead coexist as two or more phases). 

In a system containing chemical reactions, a dye is needed whose own chemical reaction matches the desired property (both in terms of kinetics and pathway). This is very difficult to achieve, in particular in situations where a system consisting of reacting bulk fluids is investigated. 

Similarly, numerous biological fluid dynamics systems exist in which the evolution of the system is affected by the presence of a dye. This could be due to interference with the metabolism of the species \cite{Sun17}. Alternatively, when studying collective motion, this could be when the organisms sense the presence of the dye, alter their behaviour and hence breakup their collective structures \cite{Fortune20}.

In this paper, we demonstrate how the concepts of dye attenuation can be extended towards tracking many normally transparent bulk liquids by using parts of the electromagnetic spectrum that do not coincide with the visible spectrum. While the use of x-rays are well known, (e.g. \cite{Lee13} and \cite{Lappan20}), here we concentrate on the longer wave lengths in the short wave infrared range of the spectrum which present much less health and safety concerns. In this range, absorption of radiation is not due to interactions with the electrons of the molecules, but instead due to the excitation of molecular vibrations.
Most substances absorb radiation within the mid-wave infrared range. This is the basis for mid-wave (MW) infrared (IR) spectroscopy and certain stand-off substance detection strategies \cite{howle2011hazardous, clewes2012stand}. Here, however, we will explore the short-wave infrared (SWIR) range, from 1050 to $\SI{2500}{\nano\metre}$. (As a reference, the visible light wavelength spectrum ranges from 380 to $\SI{700}{\nano\metre}$.) Part of the motivation for our choice is that experimental design in the mid-infrared range is cumbersome since many commonly used transparent materials (such as glass) become absorbent and thus not transparent anymore in this range. Since experimental setups involving fluids requires solid transparent objects placed in the beam path to contain the fluids, this necessitates more exotic materials to be located, increasing the complexity of the experimental setup.

In contrast, the short-wave infrared range is still absorbed in many/most fluids due to molecular vibrations but is more straight forward as standard materials such as glass remain transparent.

Recent developments in industrial machine vision \cite{AlliedVision} capitalise on SWIR imaging, such as for laser beam profiling and positioning \cite{Ophir}, non-destructive quality control of industrial processes, e.g. detecting defects in semiconductor silicon wafers \cite{Techbriefs}, food analysis and sorting \cite{Lynred}, and glass bottle inspection \cite{PhotonicsOnline} by measuring liquid levels in otherwise non-transparent vessels. Some marine, coastal and defence applications exploit the fact that the longer SWIR wavelengths can penetrate fog, smoke and other atmospheric conditions \cite{maritime}. Therefore, commercially available cameras and light sources exist. Hence, SWIR imaging appears to be a safe and promising method to target many transparent substances without adding greatly to the experimental complexity of a dye attenuation experiment using visible light.

In this paper, we demonstrate experimentally that by imaging in the short-wave infrared region, we can track the spatial and temporal evolution of depth averaged concentration fields of a fluid without the addition of dye and illustrate this for three test cases. In \S\ref{vibrations} and \S\ref{attenuationtheory} we briefly review the physico-chemical background of absorption of SWIR radiation and the fundamentals of the dye attenuation technique. We then outline the experimental configuration (\S\ref{experimentsetup}) and image analysis strategies (\S\ref{imageanalysis}) we use to illustrate the utility of the technique. 

A useful canonical case is to determine the volume (via the spatial height profile) of evaporating droplets on glass slides (\S\ref{casestudy1}), a situation which is of practical interest not only for the fluid dynamics of decontamination \cite{Landel16}, but also for microfluidic experiments \cite{Kantsler12,Hacker96}. The second example is the initial motivation of this work, the absorption of a water droplet into  a polymer (\S\ref{casestudy2}). In both these two cases, we outline appropriate calibration procedures and demonstrate the accuracy of the method against an analytical mass/weight balance. 

Finally in \S\ref{conclusion}, we draw our conclusions and outline how our method could be extended towards measuring the concentration of different substances (e.g. methyl salicylate). We  discuss the scope and potential impact of applying this technique in different areas of fluid dynamics to gain new quantitative information. 

\section{SWIR absorbance imaging}

\subsection{Vibrations in the SWIR region} \label{vibrations}
When infrared radiation interacts with matter, molecular bonds vibrate, giving rise to absorption bands \cite{Miller91}. Considering a molecular bond as a harmonic diatomic oscillator, the most fundamental and intense absorption bands correspond to the  fundamental transition from the ground vibrational $n = 0$ state to the first vibrational $n = 1$ state. For most chemical compounds, these lie in the mid-infrared region (MIR, wavelengths between $2500$ and $\SI{50000}{\nano\metre}$). 

Absorption bands visible in the short-wave infrared region correspond almost exclusively to overtones and combination transitions \cite{impopen} leading to a more complex absorption spectra with broad peaks. These bands arise from anharmonicity owing to repulsive forces between the vibrating atoms, namely the molecules do not behave like ideal oscillators. This anharmonicity allows transitions from the ground vibrational $n = 0$ state to higher vibrational $n = m > 1$ states (overtones). Furthermore, combinations of different vibrational transitions can be observed, denoted `combination transitions'. The intensity of these SWIR absorption bands also depends on anharmonicity of a particular vibration \cite{Bokobza02}. Vibrations with large anharmonicity constants have large overtone intensities. This is particularly apparent for $X$--$H$ bonds that arise from $X H_n$ functional groups since hydrogen is a small molecule. Hence, these vibrations dominate the spectra in the SWIR region. This is in contrast to what we see in MIR, where the key feature is the strength of the dipole moment rather than the anharmonicity, leading to polar bonds such as $C=O$ dominating the spectra \cite{Siesler08}.
\begin{figure}
    \centering
    \includegraphics[width=\textwidth]{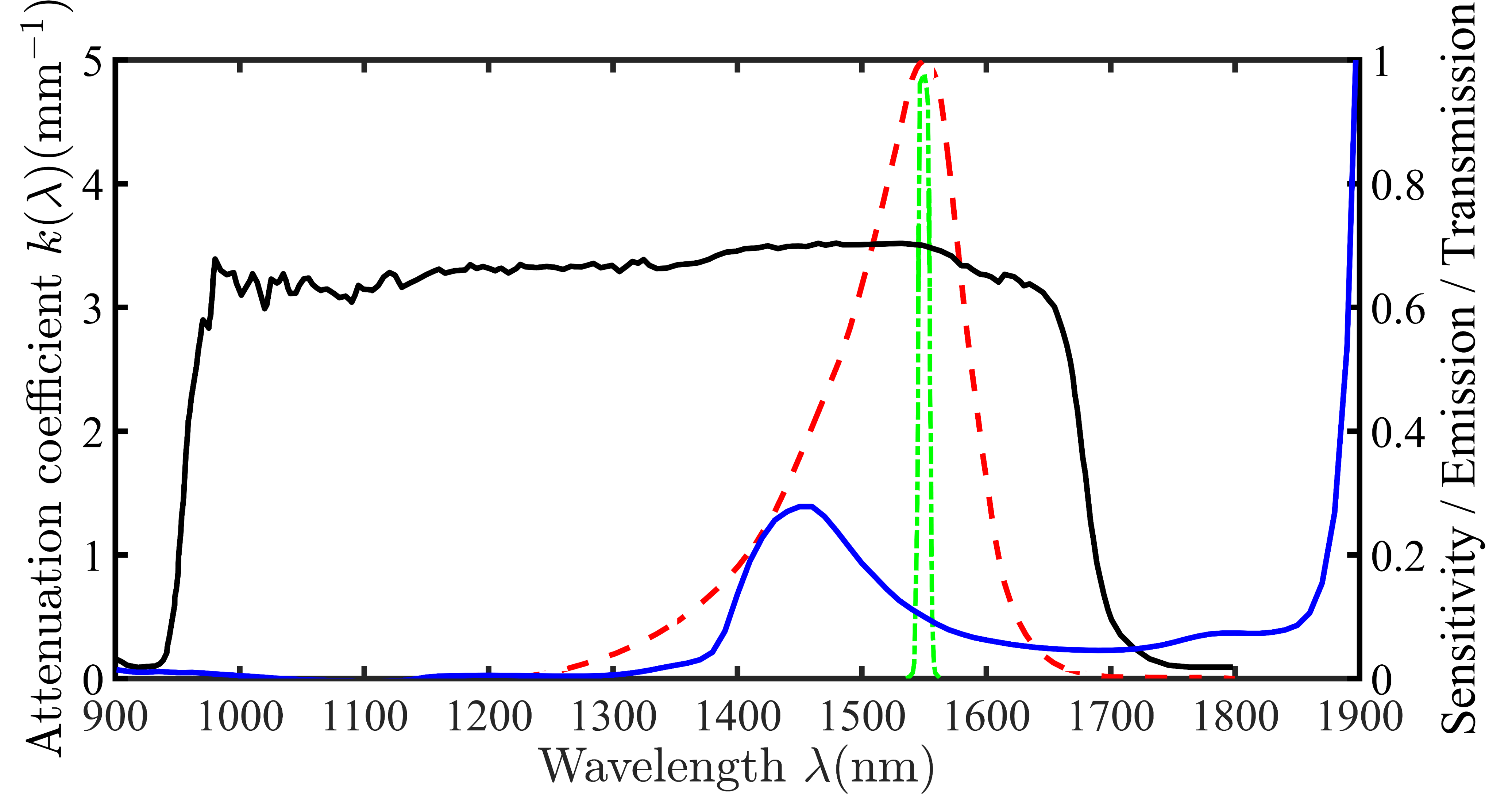}
    \caption{Absorbance plot for the development of the SWIR system, showing how the attenuation coefficient of water (blue curve), the sensitivity of the SWIR camera (black curve), the fractional emission of the IR LEDs relative to their maximum intensity (dashed red curve) and the fraction of light transmitted through the corresponding IR filter (dash-dotted green curve) vary as a function of wavelength in the short-wave infrared region. As a reference, the visible light wavelength spectrum ranges from 380 to $\SI{700}{\nano\metre}$.
    }
    \label{fig:Fortune1}
\end{figure}
Hence, for a given fluid, by choosing a suitable SWIR wavelength where the fluid strongly attenuates incident light (namely by falling in an absorbance band), we can  track fluid concentrations by measuring the light attenuation and then relating it to the fluid concentration. In the rest of this paper, for experimental convenience we use water as our main working fluid. However, the same principles apply for most other fluids. The attenuation coefficient $k$ of water as a function of wavelength $\lambda$ is plotted as a blue curve in figure \ref{fig:Fortune1}. As can be seen, water has absorbance bands (blue curve) around $\SI{1450}{\nano\metre}$ and $\SI{1930}{\nano\metre}$. The $\SI{1450}{\nano\metre}$ peak arises from the first overtone of $O$--$H$ stretching, while the \SI{1930}{\nano\metre} peak arises from a $O-H$ stretch/deformation combination \cite{Wilson15}. For comparison, we also show the spectral response curve of a typical SWIR camera with an Indium gallium arsenide sensor, which is sensitive between approximately $950$ and $\SI{1700}{\nano\metre}$ (black curve in figure \ref{fig:Fortune1}). For experimental convenience, we chose to target the right-hand flank of the absorbance peak around $\SI{1550}{\nano\metre}$ in our dye attenuation SWIR experiments.

\subsection{Light attenuation theory} \label{attenuationtheory}
The decay of intensity $I$ of electromagnetic radiation whilst passing through a single-species material is
\begin{equation}
    \frac{\partial I}{\partial z} = - \mu I, \label{eq:lambertbeer}
\end{equation}
where $\mu$ is the attenuation coefficient, assumed to be constant for water, and $z$ is the distance along a ray of light. Note that including a minus sign in the definition ensures that $\mu$ is always positive. Integration between 0 and the distance along the ray yields an expression commonly known as the Beer-Lambert law.
\begin{equation}
    A \equiv -\log\frac{I}{I_0}= \mu z, \label{eq:idealrelationship}
\end{equation}
where $A$ is the absorbance, and $I_0$ the light intensity at $z=0$. In photometric experiments, $I_0$ is determined by measuring a sample without the absorbing species. In our experiments below, we define $I_0$ to denote the true intensity of the reference image and $I$ to denote an image with the absorber sample present. However, we are not able to measure $I$ and $I_0$ directly, rather we measure the digitised pixel intensity outputted by the camera $P =\Xi(I)$ and $P_0 =\Xi(I_0)$. Note that $\Xi(I)$ and $\Xi(I_0)$ are not necessarily the identity function i.e. it is common to find small signals arising from pixels that are nominally black (e.g. when imaging black cardboard), indicating a small amount of stray radiation (or noise) that contributes to the signal of each pixel. However, we assume that the function $\Xi$ is  to good approximation a linear function. This assumption was validated through supplementary experiments where the camera was aimed at an image with uniform intensity, and then a range of different exposure times were tested.

We denote the average digitised pixel intensity across a region of black card in the image and reference image as $P_{\text{black}}$ and $P_{0, \, \text{black}}$ respectively. $I$  and $I_0$ can be related to the experimentally measurably quantities $\{ P, \, P_{\text{black}} \}$ and $\{ P_0, \, P_{\text{black}} \}$ through
\begin{equation}
I \approx P - P_{\text{black}}, \quad I_0 \approx P_0 - P_{0, \, \text{black}}.
\end{equation}
Hence, the absorbance relation defined in (\ref{eq:idealrelationship}) can be written as
\begin{equation}
    A = - \log{\left(\frac{I}{I_{0}}\right)} = - \log{\left(\frac{P - P_{\text{black}}}{P_0 - P_{0, \, \text{black}}}\right)}. \label{eq:absorbanceequation}
\end{equation}
\subsection{Experimental Configuration} \label{experimentsetup}
\begin{figure}
    \centering
    \includegraphics[width=0.5\textwidth]{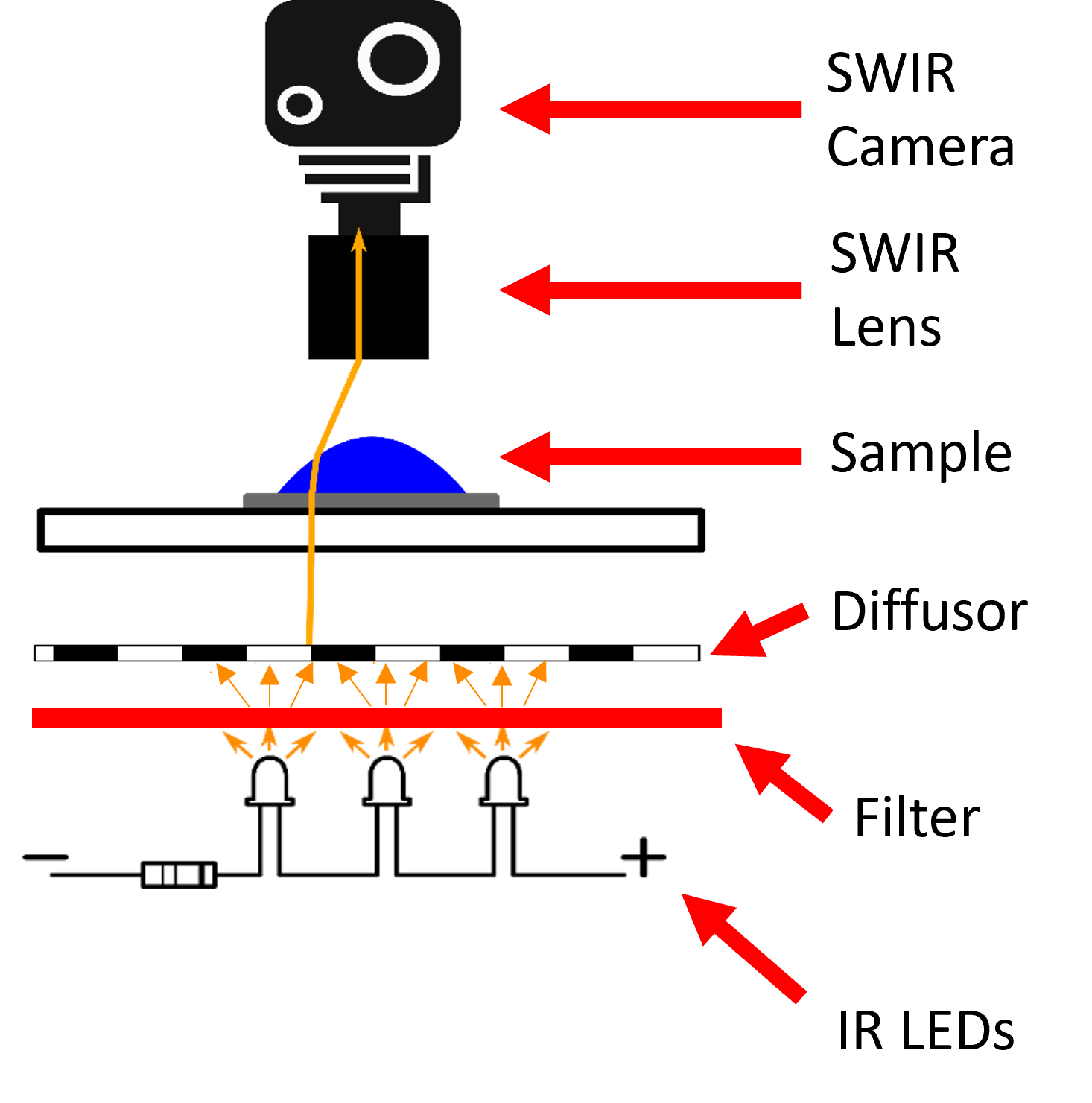}
    \caption{Short-wave infrared (SWIR) based absorbance imaging tracking water concentrations: Experimental schematic. 
    }
    \label{fig:Fortune2}
\end{figure}
Figure \ref{fig:Fortune2} gives a schematic of the experimental apparatus. A SWIR camera (WiDy SenS 640, New Imaging Technology, France) with a \SI{100}{\milli \metre} fixed focal length SWIR lens (TECHSPEC, Edmund Optics, USA) aligned with its optical axis normal to the plane of the glass slide on which the sample was placed, recorded the evolution of the samples with spatial resolution of the order of \SI{50}{\micro\metre} per pixel. The frame rate varied depending on the system imaged but could be increased to a maximum  of $120$fps.

Illumination was provided by a grid of \SI{1550}{\nano\metre} Infrared LEDS (L12509 series, Hamamatsu, Japan, emission spectra given by the dashed red curve in figure \ref{fig:Fortune1}), placed below a \SI{1550}{\nano\metre} SWIR band-pass filter (TECHSPEC, Edmund Optics, USA, transmission spectra given as the green curve in figure \ref{fig:Fortune1}). The lights were switched by an Arduino Uno (Arduino.cc, Italy) connected to a Low-Side Switch Shield (BTF3050TE; Infineon, Germany). To ensure spatial uniformity of the intensity of the light produced by the LEDs, diffusers were added to the light path just below the sample, either diffusor paper sheets or \SI{50}{\milli\metre} by \SI{50}{\milli\metre} ground glass diffusors (Edmund Optics, range of different grits).

To mitigate fluctuations in the intensity of the LEDs arising from temperature changes, the LEDs were mounted high on their legs, driven in constant current mode and cooled by a fan (see figure \ref{fig:Fortune2}(b)). Additional experiments were performed without a sample, investigating the stability of the illumination produced by the LEDs as they heated up. It was found that applying cooling from the fan achieved stability, namely the LEDs reached a constant temperature, after waiting for a period of ten minutes.
\subsection{Image Analysis} \label{imageanalysis}
Black cardboard was placed in the field of view of the camera to provide a region of the image from which $P_{\text{black}}$ and $P_{0, \, \text{black}}$ can be determined. The open source image processing package Fiji \cite{Schneider12,Schindelin12} was utilised to locate a region of interest and a black reference region in each image. For any particular experiment, these regions were taken to be in the same part of the image for each frame. 
Absorbances were obtained by computing equation (\ref{eq:absorbanceequation}) utilising MATLAB scripts that used MATLAB's Image Processing Toolbox \cite{Matlab20}. Distances measured in pixels were converted to distances in millimetres through the conversion $160.6$ pixels $\rightarrow \SI{10}{\milli\metre}$ obtained from images of a ruler placed in place of the microscope slides. No significant optical distortion was noted, being to good approximation spatially and directionally invariant.

Despite the mitigation strategies discussed above in \S\ref{experimentsetup} to limit LED intensity fluctuations, small variations, both in time and space, still occurred in pixels in the region of interest that did not contain the sample. 

This led to non-zero absorbance values in regions where we expect nothing. It was found that in systems with an inert background (e.g. flows on impermeable glass slides but not flows through polymeric hydrogel), these errors could be mitigated through the use of an additional image processing step. If the difference between the black corrected intensity of a pixel and the corresponding pixel in the reference image was less than a fixed constant $X_I$, namely
\begin{equation}
\left | \left( P   - \left\langle P_{\text{black}} \right\rangle \right) - \left( P_0 - \left\langle P_{0,\text{black}} \right\rangle \right) \right | < X_I,
\end{equation}
then we declared that this pixel was purely noise and set its absorbance to zero. For our particular configuration, we set $X_I = 100$. Since the background intensity and the black intensity were typically around $7500$ and $1000$, respectively, $X_I = 100$ corresponded to a vertical water height of $\SI{12.4}{\micro\metre}$, much lower than other uncertainties in the setup.  

\section{Case Study I: Water drop on an impermeable glass slide} \label{casestudy1}
Motivated by the experimental challenges that arise when studying microfluidic configurations \cite{Kantsler12} or channels \cite{Hacker96}, as our first case study we seek to determine the evolution of a droplet volume on a glass slide.
\subsection{Calibration} \label{casestudy1calibration}
\begin{figure}
    \centering
    \includegraphics[width=\textwidth]{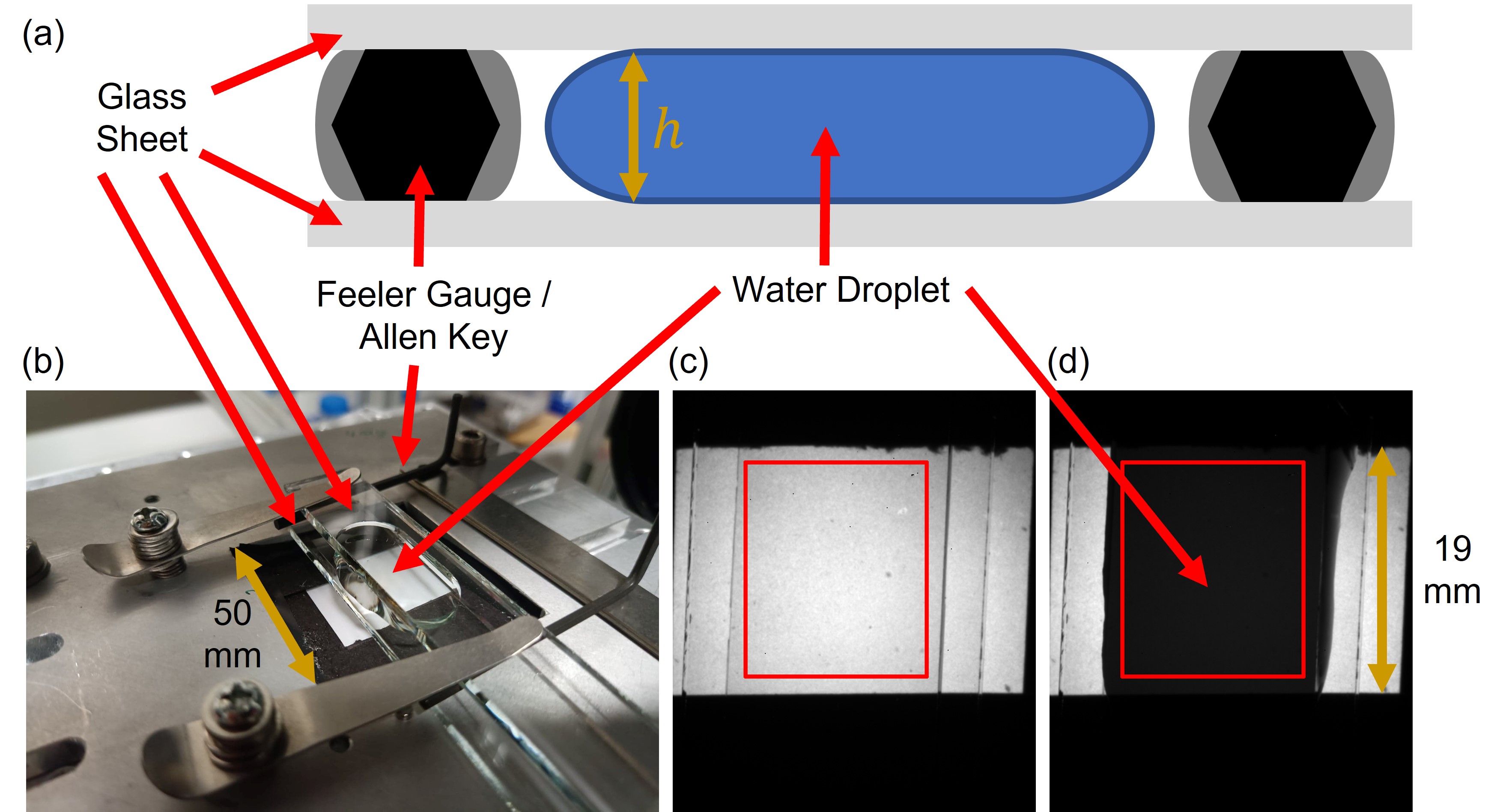}
    \caption{Case study I: experimental estimation of the effective attenuation coefficient of water $\mu_{\text{w}}$. (a) Schematic and (b) photograph of the experimental apparatus. (c,d) Photographs of an experiment with gap height $h = \SI{2}{\milli\metre}$. (c) Reference image without any water. (d) Image once the gap has been filled with water, showing the strong attenuation of the light. In both (c) and (d) the red rectangle denotes the region of interest, chosen individually for each experiment. 
    }
    \label{fig:Fortune3}
\end{figure}
The usual practice in dye attenuation experiments involving visible light is to calibrate the effective attenuation coefficient of the dye, noting that for simplicity the linear equation (\ref{eq:lambertbeer}) is assumed to hold. This effective attenuation coefficient is also a function of both the finite width of the illumination spectrum and the sensitivity of the camera used. Similarly, in our SWIR configuration, we need to determine experimentally the effective attenuation coefficient of the working fluid $\mu_{\text{w}}$ for the particular experimental apparatus. For example, in the present study where we are using water as the working fluid, changing the salinity and hardness of the water used or the temperature of the sample \cite{kakuta2009temperature} would alter $\mu_{\text{w}}$.

We determined $\mu_{\text{w}}$ by imaging a series of water-filled gaps, which were constructed as illustrated in figure \ref{fig:Fortune3}. Two \SI{3}{\milli\metre} thick glass sheets (Go Glass, Cambridge) were placed at a known fixed vertical height $h$ apart using either feeler gauges (Stainless Steel, Zhibeisai, China) or Allen keys (Facom, RS Components, UK). This gap was then filled with reverse osmosis (RO) water using a pipette. Hence, the absorbance of a height $h$ of water could then be estimated as the average absorbance of a region of the gap that was completely filled with water relative to a reference image of the setup before the water was added. Figure \ref{fig:Fortune3}(c) shows the reference image. Figure \ref{fig:Fortune3}(d) shows the image with the water sample present, clearly showing the light absorption. The gap height $h$ is $\SI{2}{\milli\metre}$. Here, the region of interest over which the absorbance was averaged is marked by a red box.

\begin{figure}
    \centering
    \includegraphics[width=\textwidth]{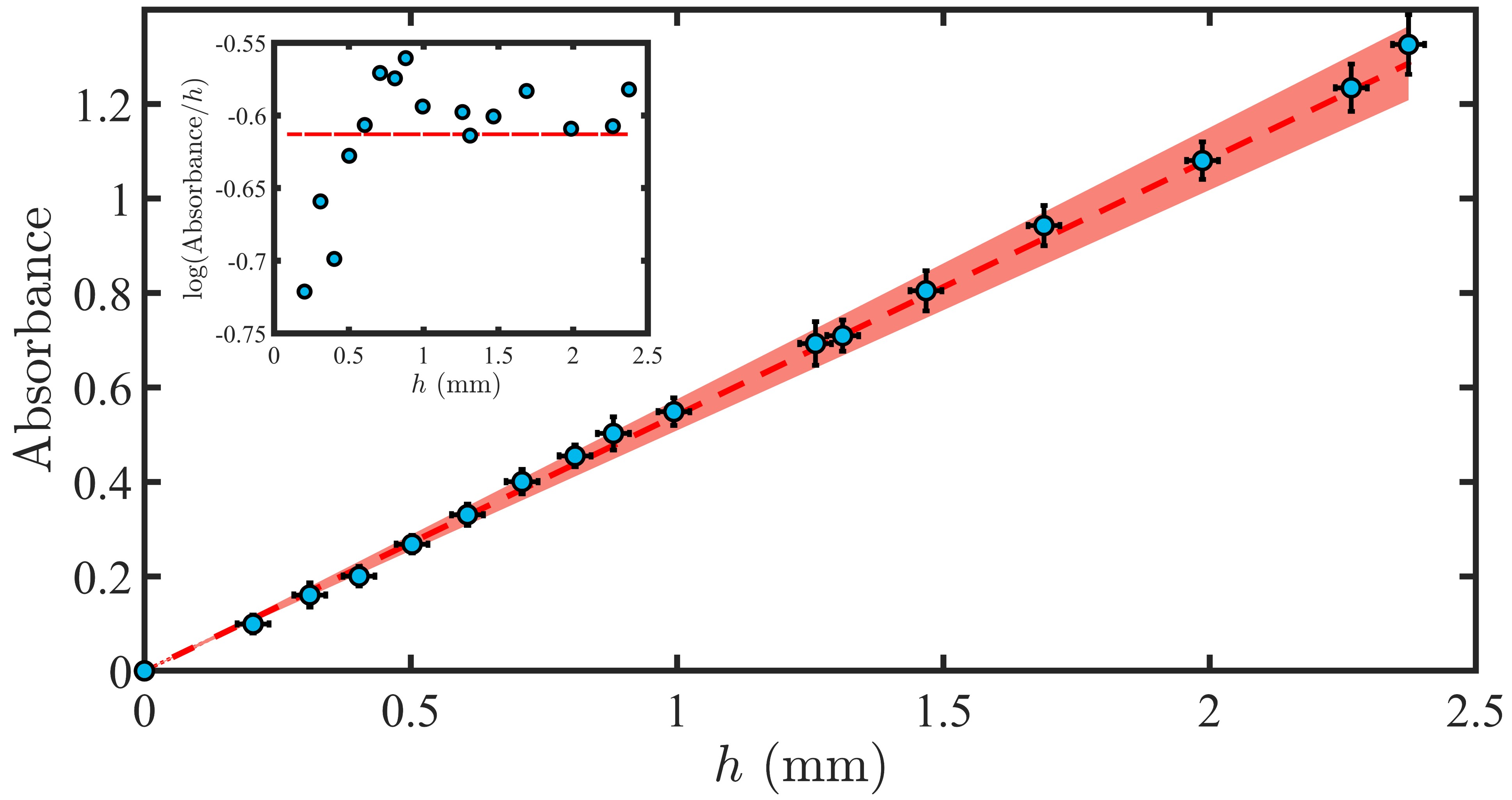}
    \caption{Plot showing how the absorbance of the water gap varies as a function of its height $h$. Blue circles (\gtf{\fullcircle}) denote distinct experiments. The corresponding black error bars were determined using $h = \SI{0.03}{\milli\metre}$ and (\ref{eq:deltaA}). The red dashed line (\sbd{\dashed}) is the multiplicative line of best fit $A = \mu_{w} h$ with corresponding error bar displayed as a light red region). Inset: $\log{(\text{Absorbance/h})}$ plotted as a function of $h$. The red dashed line (\sbd{\dashed}) denotes $\log{\left(\mu_{w} \right)}$.}
    \label{fig:Fortune4}
\end{figure}
Figure \ref{fig:Fortune4} plots in blue how the average absorbance of the water varies with water height $h$ across sixteen distinct experiments with corresponding error bars plotted in black. We have also plotted the reference zero water case, namely that $h = 0$ should lead to zero absorbance. However, we note that since there will be reflections at all glass-water interfaces, the extrapolation of the experimental data down to $h = 0$ could be expected to lead to an absorbance slightly larger than the zero water case. A seventeenth experiment with $h = \SI{0.1}{\milli\metre}$ was performed but discarded, owing to a noticeable deformation of the glass plates that led to the height profile $h$ varying spatially. This spatial variation was attributed to a combination of the strong capillary forces at the water-glass interface and height variations in the feeler gauges arising from compression of the thin metal.

The dominant error in the gap height $h$, $\Delta h$ originated in the non-uniform thickness  of the feeler gauges of the Allen keys used to separate the top and bottom glass slides. From thickness measurements utilising a micrometer at different points we estimate this error as $\Delta h \leq \SI{30}{\micro\metre}$, assumed to be constant across all experiments. The error in the absorbance $\Delta A$ arose from a combination of different experimental factors, such as the spatial and temporal fluctuations of the LEDs, possible variations in water quality or varying temperatures. Hence, a statistical estimate of the experimental error, estimated independently for each experiment, is 
\begin{equation}
    (\Delta A)_i = 2(\sigma_a)_i, \label{eq:deltaA}
\end{equation}
where $(\sigma_a)_i$ was the standard deviation of absorbance values in the region of interest for experiment $i$. Looking at figure \ref{fig:Fortune4}, the dominant contribution to $\Delta A$ has multiplicative rather than additive character, namely data points corresponding to large gap heights have larger vertical error bars. 

More formally, suppose we have $N$ experiments, each with gap height $h_i$ and average absorbance $A_i$ where $i \in [1,N]$. Utilising (\ref{eq:idealrelationship}), multiplicative character means
\begin{equation}
    A_i = \mu_{\text{w}} (1 + \epsilon_i) h_i \rightarrow \log{(A_i / h_i)} = \log{\mu_{\text{w}}} + \log{(1 + \epsilon_{i})} \approx \log{\mu_{\text{w}}} + \epsilon_i,  
\end{equation}
where $\epsilon_i$ is the uncertainty for experiment i, assumed to be normally distributed with unknown variance. Utilising the full dataset, $\mu_{\text{w}}$ can thus be estimated using
\begin{equation}
    \log{\mu_{\text{w}}} = \frac{1}{N}\sum_{i = 1}^{N} \log{\left( \frac{A_i}{h_i} \right)} \Longrightarrow \mu_{\text{w}} = 0.542. \label{eq:watermuvalue}
\end{equation}
Furthermore, upper and lower bounds for $\mu_{\text{w}}$, $\mu^{+}_{\text{w}}$ and $\mu^{-}_{\text{w}}$, can be estimated through
\begin{equation}
    \log{\mu^{+}_{\text{w}}} = \frac{1}{N}\sum_{i = 1}^{N} \log{\left( \frac{A_i + \Delta A_i / 2}{h_i - \Delta h / 2} \right)} \Longrightarrow \mu^{+}_{\text{w}} = 0.575,
 \end{equation}   
 \begin{equation}   
    \log{\mu^{-}_{\text{w}}} = \frac{1}{N}\sum_{i = 1}^{N} \log{\left( \frac{A_i - \Delta A_i /2}{h_i + \Delta h / 2} \right)} \Longrightarrow \mu^{-}_{\text{w}} = 0.509.
\end{equation}
The insert to figure \ref{fig:Fortune4} plots $\log{A_i/h_i}$ for all the experiments. As can be seen, this fit places more weight on the data points corresponding to lower $h$. From this data, we estimate to three significant figures $\mu_{\text{w}} = 0.542$, $\mu^{+}_{\text{w}} = 0.575$ and $\mu^{-}_{\text{w}} = 0.509$.

\subsection{Optical determination of droplet volume}
\begin{figure}
    \centering
    \includegraphics[width=\textwidth]{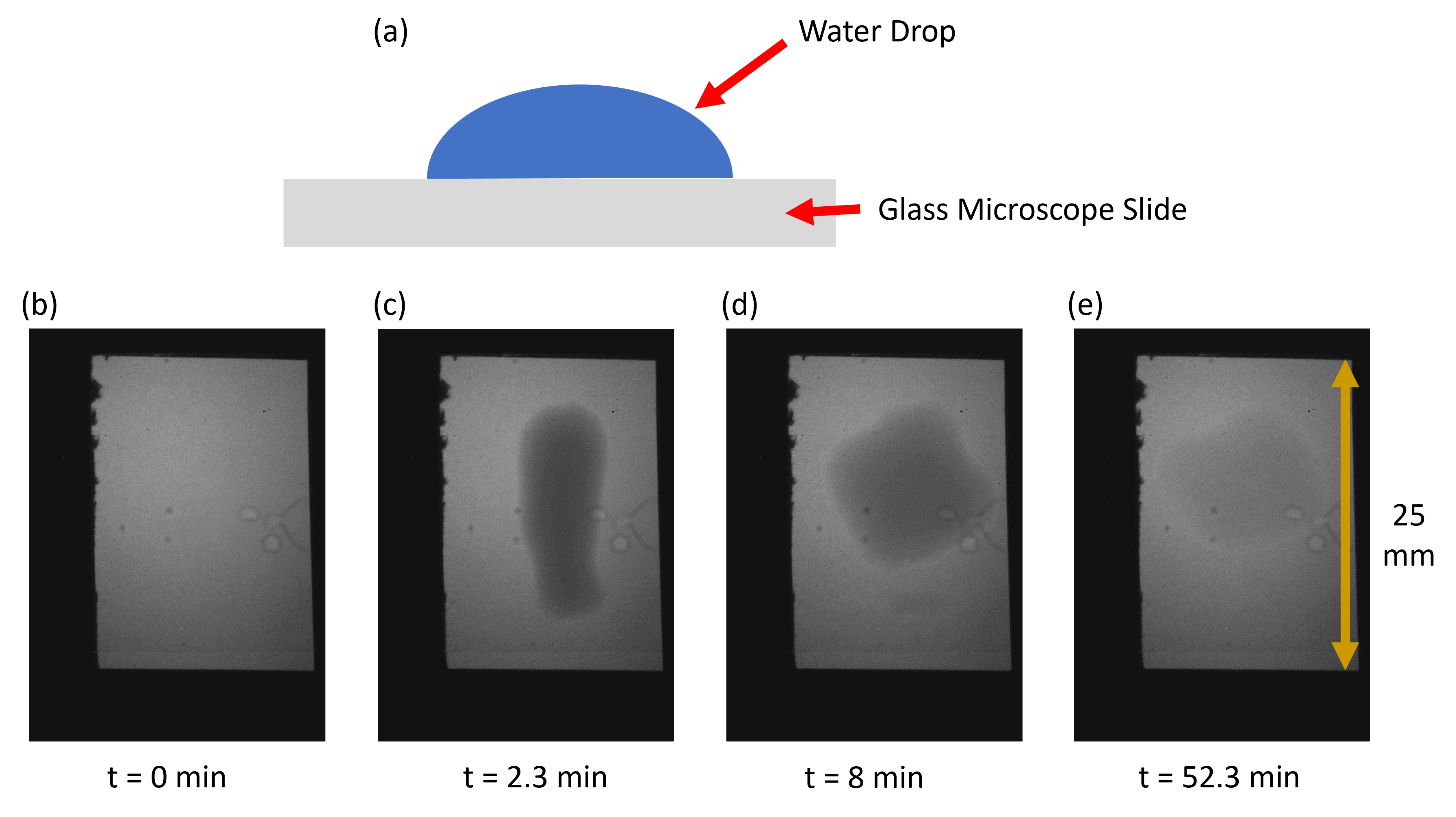}
    \caption{Water drop drying on an impermeable glass slide. (a) Schematic of the experimental configuration. (b-e) Sequence of photographs showing a typical experiment for a \SI{40}{\micro\litre} water drop. (b) Reference state. (c) Drop insertion. (d) Drop spreads. (e) Drop evaporates.
    }
    \label{fig:Fortune5}
\end{figure}
Figure \ref{fig:Fortune5}(a) presents a schematic of the experimental configuration used to measure water droplet volume on microscope slides. Note that the water drop are unconfined, namely that there is no longer a top glass slide in the optical path that was only there for calibration purposes. Starting with the initial reference state of a glass slide (figure \ref{fig:Fortune5}(b)) a droplet of water was placed using a micro-pipette (figure \ref{fig:Fortune5}(c)). This droplet spread out across the slide (figure \ref{fig:Fortune5}(d)) and eventually evaporated over time (figure \ref{fig:Fortune5}(e)). Images were taken at regular intervals (either every $\SI{20}{\second}$ or $\SI{30}{\second}$, fixed across any given experiment). Following the procedure given in \S\ref{imageanalysis}, for each image the region of interest was selected such that it fully encompassed the water droplet without the black reference. The droplet volume was estimated by summing absorbance values of all the pixels in the region of interest, converting from pixels$^2$ to $\si{\metre^2}$, and then applying (\ref{eq:idealrelationship}) together with (\ref{eq:watermuvalue}) to convert absorbance to vertical heights of water.

\begin{figure}
    \centering
    \includegraphics[width=\textwidth]{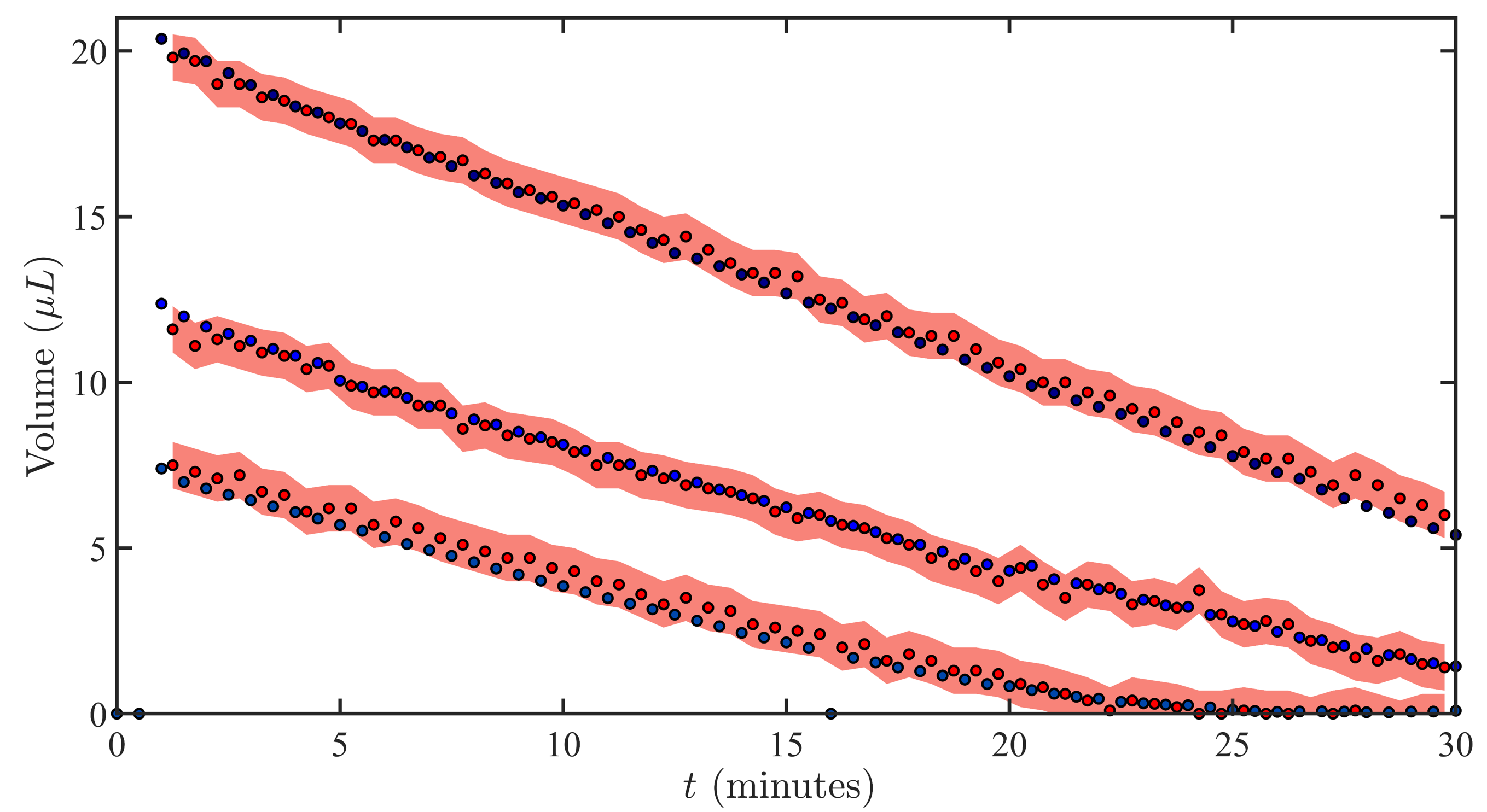}
    \caption{Temporal evolution of the water drop volume under evaporation. The blue and red dots denote measurements from the SWIR imaging and measuring balance respectively. The red shaded region denotes the uncertainty of the measuring balance ($\pm \SI{0.7}{\milli\gram}$) }
    \label{fig:Fortune6}
\end{figure}

The optically determined droplet volume was validated by removing the slide from the aparatus and weighing it on an analytical balance (PX224M Pioneer, Ohaus, USA) to obtain a comparison measurement. Figure \ref{fig:Fortune6} plots, for a range of initial drop sizes, the temporal evolution of the drop volume, obtained both by the SWIR imaging (\gtf{\fullcircle}) and by the measuring balance (\sbd{\fullcircle}). Note that different initial droplet sizes lead to different slopes due to different evaporation rates.

As can be seen, the SWIR imaging reproduces the volume of the droplet obtained by the measuring balance well within the tolerance region of the balance. Furthermore, it can be seen that the balance readings fluctuate considerably more in time than the SWIR measurements. This is particularly important for very small droplet volumes (less than \SI{1}{\micro\litre}) where this smoothness allows these volumes to be tracked.  Finally, our measurements track evaporative loss. Since in conventional dye attenuation, the amount of dye remains constant during the evaporation of the bulk phase, conventional dye attenuation can not track evaporative loss.

It is worth noting that our analysis neglects refraction effects at the edge of the droplet. This simplification is suitable since experimentally illumination is provided through diffuse rather than collimated light, minimising the dark edge to the droplets resulting from refraction.

\section{Case Study II : Water drop on an absorbent hydrogel sheet} \label{casestudy2}
In the second case study we consider the absorption, spreading and evaporation of a water droplet on a thin layer of absorbing polymer, analogue to the work of Etzold et. al. \cite{Etzold22}.  This is a more complex situation where dye cannot be used to track the fluid flow. The polymer chains in the hydrogel absorb the dye, meaning that dye concentrations no longer remain proportional to bulk fluid concentrations.

\subsection{Swelling polymer system}
The experiments in this section utilised commercially available medical-grade hydrogel pads (Hydrogel Nipple Pads, Medela, Switzerland), which were manufactured as approximately $8\times\SI{8}{\centi \metre \squared}$ sheets. Upon delivery, the hydrogel pads appear as a rubbery sheet with a slightly tacky surface. After prolonged exposure to the atmosphere over a number of hours (at ca. 25-\SI{40}{\percent} relative humidity, ca. \SI{22}{\celsius}), their initial thickness was determined with a micrometer as $a_0=\SI{1.13}{\milli \metre}\pm\SI{0.1}{\milli \metre}$. One side of a sheet was affixed to a thin plastic sheet (impermeable to water) of approximate thickness \SI{0.04}{\milli \metre} through the use of optical glue (Norland Optical Adhesive NOA 68, Edmund Optics). The other side was initially protected by an easily removable plastic film. The hydrogel sheets incorporated a sparse nonwoven gauze during their manufacture. The gauze was added probably to provide both mechanical coherence and strength in the plane of the sheet. However, this gauze did not appear to affect the swelling of the sheets or interfere with our visualisations and measurements. In our experiments, these larger sheets were cut into approximately $16\times\SI{16}{\milli \metre \squared}$ coupons. The coupons were then glued via the thin impermeable plastic sheet onto $76\times\SI{26}{\milli \metre \squared}$ microscope slides (Menzel, Germany) with UV-cured Norland Optical Adhesive NOA 68 optical glue  (Thorlabs, USA). 

Utilising Karl-Fischer titration, the water content of a typical hydrogel pad that had equilibrated with the laboratory air was measured to be $\SI{22}{\percent}$. To a good approximation, polymer swelling is driven by entropic and pairwise interactions between solvent and polymer molecules and charged groups if these are present \cite{FloryBook}. We tested for the presence of ionic groups by swelling tests in solvents which are less able to support the dissociation of the counter-ion.  We found that isopropanol and ethanol did not swell the hydrogel perceptibly and, when placed in saturated sodium chloride solutions, the swelling was drastically reduced. We take these observations as evidence of the presence of charged groups within the hydrogel. 

\subsection{Calibration} \label{casestudy2calibration}
\begin{figure}
    \centering
    \includegraphics[width=\textwidth]{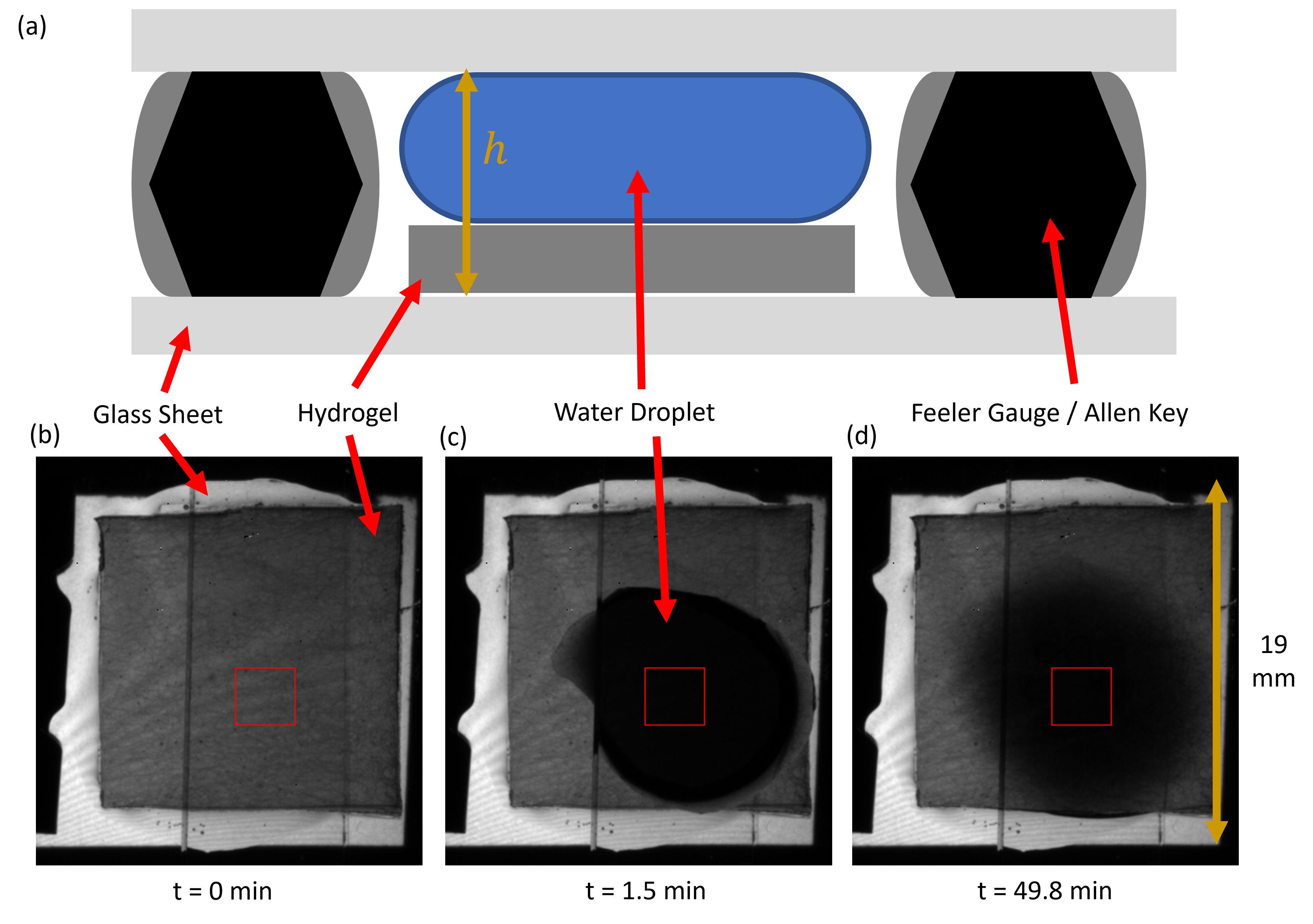}
    \caption{Experimentally estimating the attenuation coefficient of water absorbed by hydrogel $\mu_[\text{g}]$. (a) Schematic of experimental apparatus. (b-d) Experimental images for an experiment with total gap height $h = \SI{3.48}{\milli\metre}$. (b) Reference image. (c) Image when the gap has been just filled with water. (d) Image once the hydrogel has completely swelled to fill the gap. In (b), (c) and (d) the red rectangle denotes the region of interest, chosen individually for each experiment.}
    \label{fig:Fortune7}
\end{figure}
We found that the attenuation coefficient of water at our observation wavelength differed between free and absorbed water. This is because when water absorbs into a polymer, hydrogen bonds form between the water molecules and the polymeric network \cite{Gilormini18}. These bonds causes the peak arising from the first overtone of $O-H$ stretching to slightly shift to a higher wavelength, splitting into a complicated sub-band structure that is dependent on the exact polymer composition of the hydrogel \cite{Miller91}. For example, in gelatin, a new peak forms at \SI{1790}{\nano\metre} \cite{Ellis38}. This will result in a high attenuation coefficient at $\SI{1550}{\nano\metre}$ and thus a higher value for $\mu$, namely $\mu_{\text{g}} > \mu_{\text{w}}$. 

Since the hydrogel does not swell uniformly, we cannot simply swell the hydrogel to a certain water content and deduce the thickness from these measurements. Thus, we devised a refined calibration procedure which enables us to control the thickness of the hydrogel as  illustrated in figure \ref{fig:Fortune7}(a). Similar to before, two $\SI{3}{\milli\metre}$ thick glass sheets were placed a known fixed vertical height $h$ apart using a combination of feeler gauges or Allen keys. A hydrogel coupon of height $h_{\text{g}}$ glued to a microscope slide was placed in the gap and the remaining vertical space was filled with water. A region of interest was chosen that only contained water and hydrogel during the whole experiment, namely no air gaps. The glass sheets were placed sufficiently far apart to ensure that at the start of an experiment the water layer extended spatially across the whole coupon.

\begin{figure}
    \centering
    \includegraphics[width=\textwidth]{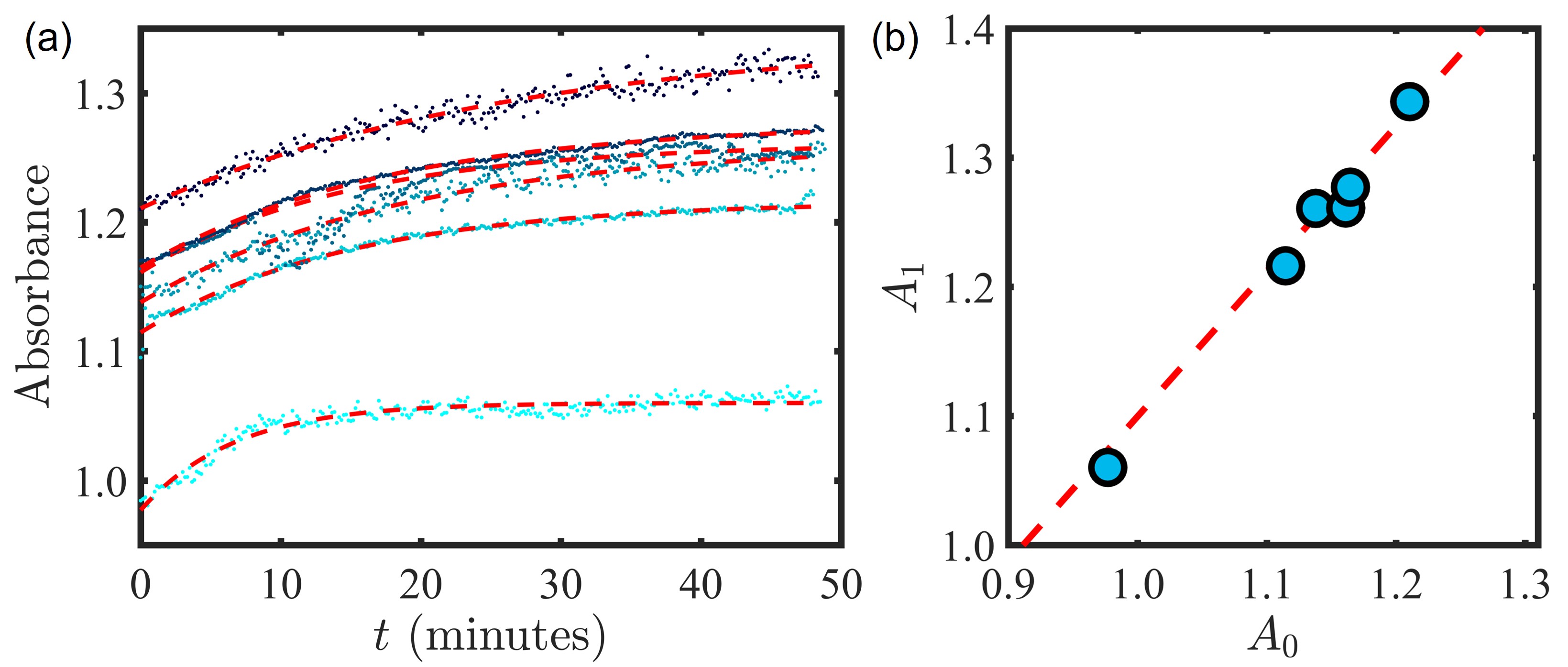}
    \caption{Calibration gap experiments with hydrogel. (a) Blue curves plots the temporal evolution of the average absorbance of the region of interest for a range of different gaps. Darker blue curves denote experiments with larger gap widths. Dashed red lines are best fit curves, obtained by fitting a decaying exponential. (b) Plot of $A_1$ \ref{eq:definingA1} against $A_0$ \ref{eq:definingA0} for all six experiments. 
    }
    \label{fig:Fortune8}
\end{figure}

The average absorbance of this region, relative to a reference image before the water was added (figure \ref{fig:Fortune7}(b)), was tracked as a function of time, taking images every ten seconds for fifty minutes. At the start of an experiment (e.g. figure \ref{fig:Fortune7}(c)), almost all the added water is `free' as it has not been absorbed yet by the hydrogel, leading to an average absorbance $A_0$ which is approximately
\begin{equation}
    A_0 = \mu_{\text{w}}\left( h - h_{\text{g}} \right), \label{eq:definingA0}
\end{equation}
where as defined above $h_g$ is the height of the hydrogel coupon. Towards the end of an experiment (e.g. figure \ref{fig:Fortune7}(d)), the hydrogel has swollen and has absorbed all the added water, leading to an average absorbance $A_1$ which satisfies
\begin{equation}
A_1 = \mu_{\text{g}}\left( h - h_{\text{g}} \right). \label{eq:definingA1}
\end{equation}
Since $h$ and $h_{\text{g}}$ can be experimentally measured using a micrometer, $\mu_{\text{g}}$ could be measured by just rearranging (\ref{eq:definingA1}). However, the large error bars on the measurements for both $h_{\text{g}}$ (due to the softness of the hydrogel) and $h$ (due to the gap is formed from combining multiple feeler gauges and Allen keys, each with their own spatial variation) lead to a large uncertainty on the resulting value for $\mu_{\text{g}}$. Hence, instead we combine (\ref{eq:definingA0}) and (\ref{eq:definingA1}) for each $h$ to give
\begin{equation}
   \mu_{\text{g}} =  \frac{A_1}{A_0}\mu_{\text{w}}. \label{eq:gelmuvalue}
\end{equation}
This is a more accurate way to determine $\mu_{\text{g}}$. Some additive errors in the computation of $A_0$ are also in $A_1$, namely the computed values for $A_0$ and $A_1$ are not independent. Hence, the resulting additive error in $A_1 / A_0$ will be lower than the sum of the errors in $A_0$ and $A_1$. Figure \ref{fig:Fortune8}(a) plots the average absorbance of the region of interest as a function of time for a range of different gap heights across six experiments. Darker blue curves denote experiments with larger gap heights. Curves of best fit are generated using the inbuilt MATLAB function fit, fitting through a non-linear least square method with three fitting parameters $(a,c,d)$ a decaying exponential curve of the form 
\begin{equation}
a + c \, e^{d t}. 
\end{equation}
Then $A_0$ and $A_1$ could be computed from these fitting parameters using the relations $A_0 = a + c$ and $A_1 = a$. Figure \ref{fig:Fortune8}(b) plots the obtained values from fitting for $A_0$ and $A_1$ using all six experiments. Note that they collapse very well using the methods of least squares onto a straight line constrained to pass through the origin (utilising the MATLAB function polyfix \cite{mjaavatten}), leading to 
\begin{equation}
   \frac{A_1}{A_0} = 1.096 \rightarrow \mu_{\text{g}} = 0.594 \, (3sf). \label{eq:calculatingmug} 
\end{equation}

\subsection{Optical determination of absorbed volume}
\begin{figure}
    \centering
    \includegraphics[width=\textwidth]{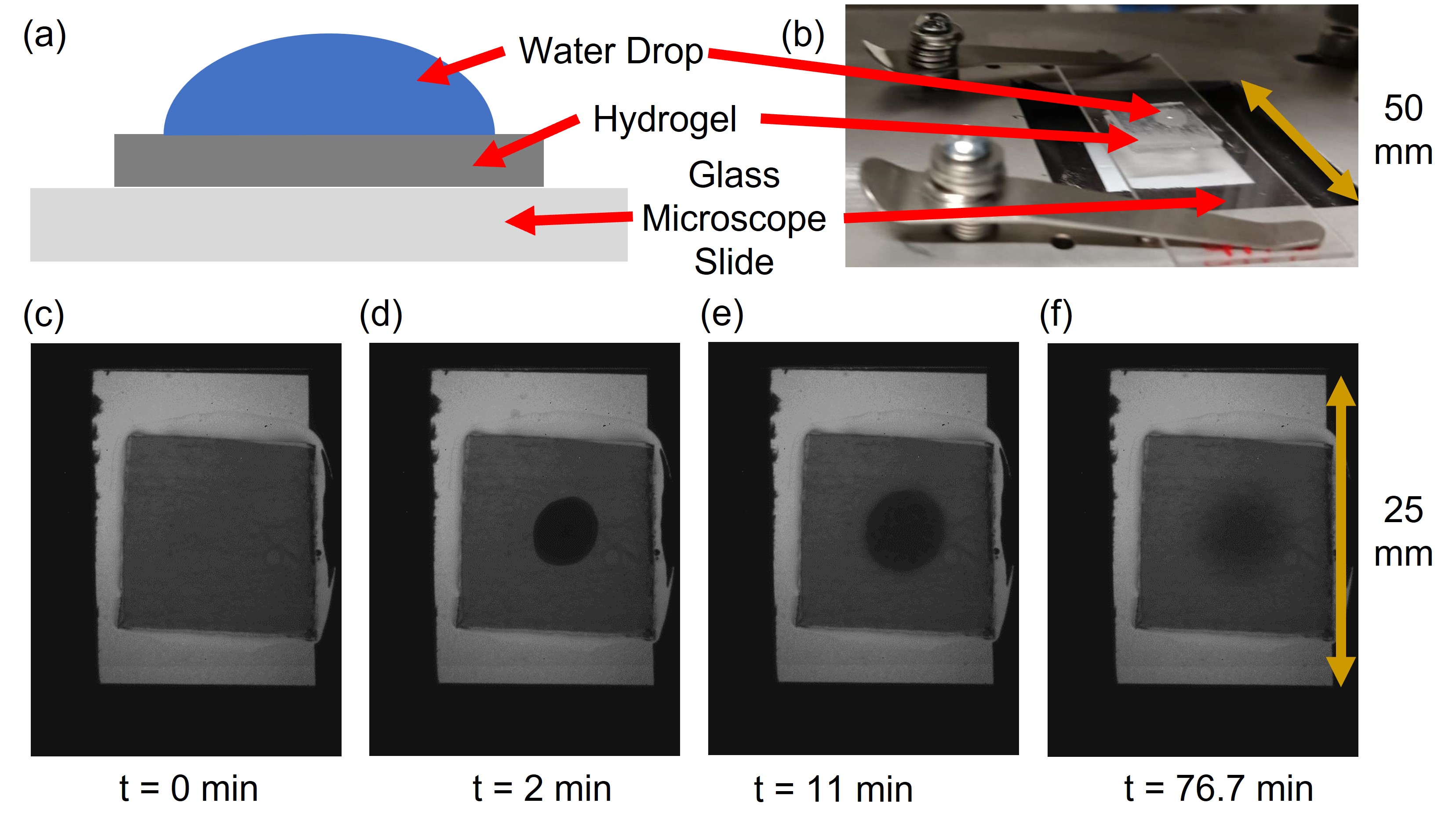}
    \caption{Water drop drying on an thin hydrogel sheet. (a) Schematic and (b) Photograph of the experimental apparatus. (c-f) Sequence of photographs showing a typical experiment for a \SI{40}{\micro\litre} water drop. (c) Reference state. (d) Drop insertion forming a blister. (e) Early time dynamics as the blister spreads. (f) Blister shrinks due to evaporation.}
    \label{fig:Fortune9}
\end{figure}
Figures \ref{fig:Fortune9}(a) and \ref{fig:Fortune9}(b) present a schematic and a photograph of what was placed in the sample region of the SWIR imaging apparatus given in figure \ref{fig:Fortune2} so that the evolution of a water drop on a hydrogel sheet could be recorded. Starting with the initial reference state of a hydrogel sheet glued onto a glass microscope slide (figure \ref{fig:Fortune9}(c)), a droplet of water was placed using a micro-pipette. This droplet is quickly absorbed by the hydrogel to form a blister (figure \ref{fig:Fortune9}(d)). This blister spreads out (figure \ref{fig:Fortune9}(e)) before diminishing due to the evaporation the water absorbed in the hydrogel (figure \ref{fig:Fortune9}(d)). Images were taken at regular intervals (either every $\SI{20}{\second}$ or $\SI{30}{\second}$, fixed across any given experiment). Following the procedure given in \S\ref{imageanalysis}, for each image the region of interest was taken to fully encompass the blister but lie fully inside the hydrogel. The blister volume was estimated by summing absorbance values of all the pixels in the region of interest, converting from pixels$^2$ to $\si{\metre^2}$, and then applying (\ref{eq:calculatingmug}) and (\ref{eq:idealrelationship}) to convert absorbances to vertical heights of water.

\begin{figure}
    \centering
    \includegraphics[width=\textwidth]{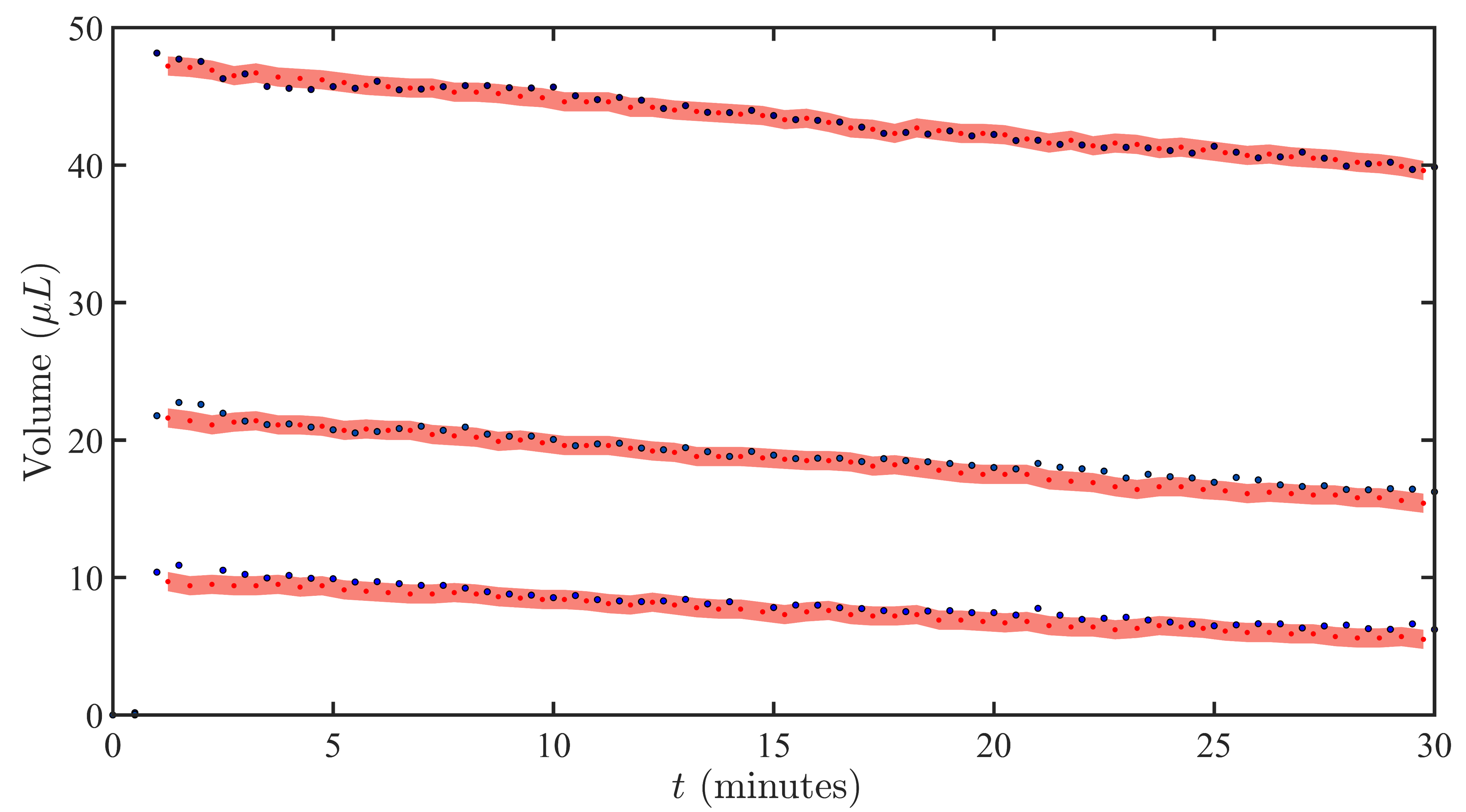}
    \caption{Temporal evolution of the blister volume under evaporation. Blue / red dots are measurements from the SWIR imaging / measuring balance respectively. The red shaded region denotes the uncertainty of the measuring balance (estimated to be $\pm \SI{0.7}{\milli\gram}$)}
    \label{fig:Fortune10}
\end{figure}

Between images, the sample was removed from the apparatus and weighed using a measuring balance (PX224M Pioneer, Ohaus, USA) to obtain a comparison measurement. Figure \ref{fig:Fortune10} plots, for a range of initial drop sizes, the temporal evolution of the blister volume, obtained both by the SWIR imaging (\gtf{\fullcircle}) and by the measuring balance (\sbd{\fullcircle}). As can be seen, the SWIR imaging reproduces the volume of the blister obtained by the measuring balance well within the tolerance of the balance for a wide range of droplet volumes.

\section{Conclusion}\label{conclusion}
In conclusion, we have presented an experimental setup that, working in the short-wave infrared region, allows fluid thicknesses to be measured optically without the need to add dye to the system for heights down to \SI{0.2}{\milli\metre} with spatial and temporal resolutions of the order of \SI{50}{\micro\metre} per pixel and $120$ fps respectively. This overcomes a inherent difficulty in many experimental systems where conventional dye attenuation methods do not work. For example, in many living biological systems, the addition of the dye induces physiological responses that lead to the system exhibiting different behaviour. Furthermore, in systems with strong chemical interactions, the dye does not follow the bulk fluid, i.e. dye concentrations no longer become proportional to bulk fluid concentrations.

Two case studies, namely considering a water drop spreading on a glass slide and a water drop spreading into a hydrogel sheet, demonstrate the scope and accuracy of this technique, in particular the capability to track very small drop volumes and very thin layers of fluid at high frame rates. Whilst SWIR spectroscopy is a rich and broad field\cite{Hindle08}, we are not aware of similar work using SWIR imaging of otherwise transparent substances to measure concentration profiles. However, we note that SWIR imaging has also been used to measure depth-averaged temperature profiles, which might be of interest for fluid experimentation as well \cite{kakuta2009temperature}.

Whilst we demonstrated the technique for water, many other common substances in fluid dynamical experiments have strong absorption bands in the SWIR region (e.g. a single sharp peak in methanol at \SI{1405}{\nano\metre} \cite{Bec16}, a sharp double peak in ethanol at \SI{1407}{\nano\metre} and \SI{1412}{\nano\metre} \cite{Bec16}, a sharp increase in glycerol around \SI{1400}{\nano\metre} followed by a gradual decrease before another sharp increase at \SI{2030}{\nano\metre} \cite{Zhou19}, a peak in acetone at \SI{1650}{\nano\metre} \cite{Wang16} and a broad peak in methyl salicylate around \SI{1862}{\nano \metre}) \cite{Xu22}. 

This will allow the determination of concentration fields in a dye-attenuation style experiments for previously inaccessible systems. This will enable new experiments in biological, chemically-driven and reacting flow systems. For example, experiments studying the fluid dynamics of decontamination could be conducted with chemically more realistic systems, which might be a disruptive technology for this field since solubility and reactivity can be better taken into account. This requires the selection of an alternative light source with a very narrow emission spectrum such as a laser.

A natural extension to this work is to modify the experimental setup to deal with reflected SWIR radiation from a sample rather than transmission. This would greatly increase the range of different system that could be probed e.g. flows through fabric. However, due to the number of different reflections and refractions that would be produced by a complex sample with multiple different interfaces, quantitative measurements would be more challenging.

Furthermore, we plan to move from illuminating with a single wavelength to sampling across multiple wavelengths, e.g. utilising a tuneable laser. This opens the door not only for quantitatively tracking of multiphase flows with three or more different fluids, but also for considering the presence of chemical reactions between them.

\section*{Acknowledgments}
GTF, MAE, JRL and SBD acknowledge funding from the Defence Science and Technology Laboratory (Dstl)

\bibliographystyle{iopart-num}
\bibliography{Fortune_et_al}

\providecommand{\newblock}{}
\begin{thebibliography}{10}
\expandafter\ifx\csname url\endcsname\relax
  \def\url#1{{\tt #1}}\fi
\expandafter\ifx\csname urlprefix\endcsname\relax\def\urlprefix{URL }\fi
\providecommand{\eprint}[2][]{\url{#2}}

\bibitem{Cenedese98}
Cenedese C and Dalziel S~B 1998 {\em Proc. Int. Symp. Flow Vis.\/} {\bf 8} 1--37

\bibitem{landel12}
Landel J~R, Caulfield C~P and Woods A~W 2012 {\em J. Fluid Mech.\/} {\bf 711} 212–258 \href{https://doi.org/10.1017/jfm.2012.388}{doi : 10.1017/jfm.2012.388}

\bibitem{davies_wykes_dalziel_2014}
Wykes M~S~D and Dalziel S~B 2014 {\em J. Fluid Mech.\/} {\bf 756} 1027–1057 \href{https://doi.org/10.1017/jfm.2014.308}{doi : 10.1017/jfm.2014.308}

\bibitem{Allgayer12}
Allgayer D~M and Hunt G~R 2012 {\em Exp. Therm. Fluid Sci.\/} {\bf 38} 257--261 \href{https://doi.org/10.1016/j.expthermflusci.2011.10.009}{doi : 10.1016/j.expthermflusci.2011.10.009}

\bibitem{Holford96}
Holford J~M and Dalziel S~B 1996 {\em Appl. Sci. Res.\/} {\bf 56} 191--207 \href{https://doi.org/10.1007/BF02249381}{doi : 10.1007/BF02249381}

\bibitem{Hacker96}
Hacker J, Linden P~F and Dalziel S~B 1996 {\em Dynam. Atmos. Oceans\/} {\bf 24} 183--185 \href{https://doi.org/10.1016/0377-0265(95)00443-2}{doi : 10.1016/0377-0265(95)00443-2}

\bibitem{Landel16}
Landel J~R, Thomas A~L, McEvoy H and Dalziel S~B 2016 {\em J. Fluid Mech.\/} {\bf 789} 630--668 \href{https://doi.org/10.1017/jfm.2015.742}{doi : 10.1017/jfm.2015.742}

\bibitem{Kim03}
Kim S and Cho Y~I 2003 {\em J. Non-Newtonian Fluid Mech.\/} {\bf 111} 63--68 \href{https://doi.org/10.1016/S0377-0257(03)00009-0}{doi : 10.1016/S0377-0257(03)00009-0}

\bibitem{Zhang19}
Zhang K, Xie R, Fang K, Chen W, Shi Z and Ren Y 2019 {\em J Mol. Liq.\/} {\bf 287} 1--7 \href{https://doi.org/10.1016/j.molliq.2019.110932}{doi : 10.1016/j.molliq.2019.110932}

\bibitem{Bertrand16}
Bertrand T, Peixinho J, Mukhopadhyay S and MacMinn C~W 2016 {\em Phys. Rev. Appl.\/} {\bf 6} 064010 \href{https://doi.org/10.1103/PhysRevApplied.6.064010}{doi: 10.1103/PhysRevApplied.6.064010}

\bibitem{Engelsberg:2013}
Engelsberg M and Jr W~B 2013 {\em Phys. Rev. E\/} {\bf 88} 062602 \href{https://doi.org/10.1103/PhysRevE.88.062602}{doi: 10.1103/PhysRevE.88.062602}

\bibitem{Etzold21}
Etzold M~A, Linden P~L and Worster M~G 2021 {\em J. Fluid Mech.\/} {\bf 925} A8 \href{https://doi.org/10.1017/jfm.2021.608}{doi: 10.1017/jfm.2021.608}

\bibitem{Hennessy21}
Hennessy M~G, M\"unch A and Wagner B 2020 {\em Phys. Rev. E\/} {\bf 101} 032501 \href{https://doi.org/10.1103/PhysRevE.101.032501}{doi : 10.1103/PhysRevE.101.032501}

\bibitem{Butler22}
Butler M and Montenegro-Johnson T 2022 {\em J. Fluid Mech.\/} {\bf 947} 1--34 \href{https://doi.org/10.1017/jfm.2022.641}{doi : 10.1017/jfm.2022.641}

\bibitem{Fortune21}
Fortune G~T, Oliveira N~M and Goldstein R~E 2022 {\em Phys. Rev. Lett.\/} {\bf 128} 178102 \href{https://doi.org/10.1103/PhysRevLett.128.178102}{doi : 10.1103/PhysRevLett.128.178102}

\bibitem{Etzold22}
Etzold M~A, Fortune G~T, Landel J~R and Dalziel S~B 2022 {\em arXiv.\/} \href{https://doi.org/10.48550/arXiv.2202.10389}{:2202.10389}

\bibitem{Doi:2009}
Doi M 2009 {\em J. Phys. Soc. Jpn.\/} {\bf 78} 052001 \href{https://doi.org/10.1143/JPSJ.78.052001}{doi: 10.1143/JPSJ.78.052001}

\bibitem{Sun17}
Sun J, Jin J, Beger R~D, Cerniglia C~E and Chen H 2017 {\em J. Ind. Microbiol. Biot.\/} {\bf 44} 1471--1481 \href{https://doi.org/10.1007/s10295-017-1970-8}{doi : 10.1007/s10295-017-1970-8}

\bibitem{Fortune20}
Fortune G~T, Worley A, Sendova-Franks A~B, Franks N~R, Leptos K~C, Lauga E and Goldstein R~E 2021 {\em J. Fluid. Mech.\/} {\bf 914} A20 \href{https://doi.org/10.1017/jfm.2020.1112}{doi : 10.1017/jfm.2020.1112}

\bibitem{Lee13}
Lee J~S, Weon B~M and Je J~H 2013 {\em J. Phys. D Appl. Phys.\/} {\bf 46} 494006 \href{https://doi.org/10.1088/0022-3727/46/49/494006}{doi : 10.1088/0022-3727/46/49/494006}

\bibitem{Lappan20}
Lappan T, Franz A, Schwab H, K\"uhn U, Eckert S, Eckert K and Heitkam S 2020 {\em Soft Matter\/} {\bf 16} 2093--2103 \href{https://doi.org/10.1039/C9SM02140J}{doi : 10.1039/C9SM02140J}

\bibitem{howle2011hazardous}
Howle C~R, Frisby A, McIntosh A~J~S, Stothard D~J~M, Dunn M~H, Robertson G, Miller W, Malcolm G and Maker G 2011 Hazardous liquid detection by active hyperspectral imaging {\em Optics and Photonics for Counterterrorism and Crime Fighting VII; Optical Materials in Defence Systems Technology VIII; and Quantum-Physics-based Information Security\/} vol 8189 (SPIE) pp 124--131 \href{https://doi.org/10.1117/12.898516}{doi : 10.1117/12.898516}

\bibitem{clewes2012stand}
Clewes R~J, Howle C~R, Stothard D~J~M, Dunn M~H, Robertson G, Miller W, Malcolm G, Maker G, Cox R, Williams B and Russell M 2012 Stand-off spectroscopy for the detection of chemical warfare agents {\em Optics and Photonics for Counterterrorism, Crime Fighting, and Defence VIII\/} vol 8546 (SPIE) pp 288--295 \href{https://doi.org/10.1117/12.974574}{doi : 10.1117/12.974574}

\bibitem{AlliedVision}
Vision A White paper: Short-wave infrared ({SWIR}) cameras offer versatile application fields in machine vision

\bibitem{Ophir}
Rieley D Beam profiling in the {SWIR} range: What you need to know \href{https://www.ophiropt.com/laser--measurement/knowledge-center/article/11570}{mks Ophir} Retrieved 2022-08-22

\bibitem{Techbriefs}
Briefs T {InGaAs} {SWIR} imagers optimize semiconductor inspection \href{https://www.techbriefs.com/component/content/article/tb/supplements/ptb/features/applications/11646}{Retrieved 2022-08-22}

\bibitem{Lynred}
Lynred Three reasons for using {SWIR} infrared imaging to improve quality control in the food and beverage industry \href{https://lynred.com/blog/three-reasons-using-swir-infrared-imaging-improve-quality-control-food-and-beverage-industry}{Lynred blog} Retrieved 2022-08-22

\bibitem{PhotonicsOnline}
Unlimited S {SWIR} for hot-end glass bottle defect inspection and imaging \href{https://www.photonicsonline.com/doc/hot-end-glass-bottle-defect-inspection-and-imaging-0001}{Photonics Online} Retrieved 2022-08-22

\bibitem{maritime}
Control S~S~V~~M Maritime \& coastal surveillance \href{https://silentsentinel.com/wp-content/uploads/2020/06/Maritime-Brochure-Silent-Sentinel.pdf}{Retrieved 2022-08-22}

\bibitem{Kantsler12}
Kantsler V and Goldstein R~E 2012 {\em Phys. Rev. Lett.\/} {\bf 108} 038103 \href{https://doi.org/10.1103/PhysRevLett.108.038103}{doi : 10.1103/PhysRevLett.108.038103}

\bibitem{Miller91}
Miller C~E 1991 {\em Appl. Spectrosc. Rev.\/} {\bf 26} 277--339 \href{https://doi.org/10.1080/05704929108050883}{doi : 10.1080/05704929108050883}

\bibitem{impopen}
Davies A~M~C An introduction to near infrared ({NIR}) spectroscopy \href{https://www.impopen.com/introduction-near-infrared-nir-spectroscopy}{IM Publications Open} Retrieved 2022-08-04

\bibitem{Bokobza02}
Bokobza L 2002 Origin of near-infrared absorption bands {\em Near-Infrared Spectroscopy\/} ed Siesler H~W, Ozaki Y, Kawata S and Heise H~M (Wiley-VCH) pp 11--42

\bibitem{Siesler08}
Siesler H~W 2008 Basic primciples of near-infrared spectroscopy {\em Handbook of Near-Infrared Analysis, Third Edition\/} ed Burns D~A and Ciurczak E~W (CRC Press (Taylor and Francis Group)) pp 7--19

\bibitem{Wilson15}
Wilson R~H, Nadeau K~P, Jaworski F~B, Trombreg B~J and Durkin A~J 2015 {\em J. Biomed. Opt.\/} {\bf 20} 030901 \href{https://doi.org/10.1117/1.JBO.20.3.030901}{doi : 10.1117/1.JBO.20.3.030901}

\bibitem{Schneider12}
Schneider C~A, Rasband W~S and Eliceiri K~W 2012 {\em Nature methods\/} {\bf 9} 671--675

\bibitem{Schindelin12}
Schindelin J, Arganda-Carreras I, Frise E, Kaynig V, Longair M, Pietzsch T, Preibisch S, Rueden C, Saalfeld S, Schmid B, Tinevez J~Y, White D~J, Hartenstein V, Eliceiri K, Tomancak P and Cardona A 2012 {\em Nature Methods\/} {\bf 9} 676--682 \href{https://doi.org/10.1038/nmeth.2019}{doi: 10.1038/nmeth.2019}

\bibitem{Matlab20}
The~MathWorks I 2020 {\em {MATLAB} and Image Processing Toolbox Release 2020b\/} Natick, Massachusetts, United States

\bibitem{kakuta2009temperature}
Kakuta N, Kondo K, Ozaki A, Arimoto H and Yamada Y 2009 {\em Int. J. Heat Mass Tran.\/} {\bf 52} 4221--4228 \href{https://doi.org/10.1016/j.ijheatmasstransfer.2009.04.024}{doi : 10.1016/j.ijheatmasstransfer.2009.04.024}

\bibitem{FloryBook}
Flory P~J 1953 {\em Principles of polymer chemistry\/} (Cornell University Press)

\bibitem{Gilormini18}
Gilormini P and Verdu J 2018 {\em Polymer\/} {\bf 142} 164--169 \href{https://doi.org/doi.org/10.1016/j.polymer.2018.03.033}{doi : doi.org/10.1016/j.polymer.2018.03.033}

\bibitem{Ellis38}
Ellis J~W and Bath J 1938 {\em J. Chem. Phys.\/} {\bf 6} 723--729 \href{https://doi.org/10.1063/1.1750157}{doi : 10.1063/1.1750157}

\bibitem{mjaavatten}
Mjaavatten A polyfix \href{https://www.mathworks.com/matlabcentral/fileexchange/54207-polyfix-x-y-n-xfix-yfix-xder-dydx}{MATLAB Central File Exchange} Retrieved 2022-08-21

\bibitem{Hindle08}
Hindle P~H 2008 Historical development {\em Handbook of Near-Infrared Analysis, Third Edition\/} ed Burns D~A and Ciurczak E~W (CRC Press (Taylor and Francis Group)) pp 3--6

\bibitem{Bec16}
Be\'{c} K~B, Futami Y, W\'{o}jcik M~J and Ozaki Y 2016 {\em Phys. Chem. Chem. Phys.\/} {\bf 18} 13666 \href{https://doi.org/10.1039/c6cp00924g}{doi : 10.1039/c6cp00924g}

\bibitem{Zhou19}
Zhou W, Liu H, Xu Q, Li P, Zhao L and Gao H 2019 {\em Spectrochim. Acta A\/} {\bf 228} 117824 \href{https://doi.org/10.1016/j.saa.2019.117824}{doi : 10.1016/j.saa.2019.117824}

\bibitem{Wang16}
Wang C, Scherrer S~T and Hossain D 2016 {\em Appl. Spectrosc.\/} {\bf 58} 784--791 \href{https://doi.org/10.1366/0003702041389193}{doi : 10.1366/0003702041389193}

\bibitem{Xu22}
Xu T 2022 {\em AIP Conf. Proc.\/} {\bf 2589} 020014 \href{https:/doi.org/10.1063/5.0112957}{doi : 10.1063/5.0112957}

\end{thebibliography}
\end{document}